\documentclass[12pt,fleqn, amsfonts]{article} \textheight=9in
\newcommand{\ra}{\rightarrow}
 
\newcommand{\be}{\begin{equation}}
\newcommand{\ee}{\end{equation}}
\newcommand{\bea}{\begin{eqnarray}}
\newcommand{\eea}{\end{eqnarray}}

\newcommand{\R}{\mathbb R}

\newcommand{\ffi}{\varphi}

\newcommand{\ep}{\quad {\vrule height 10pt width 10pt depth 2pt}}

\newcommand{\grintl}{[\kern-.18em [}
\newcommand{\grintr}{]\kern-.18em ]}
\newcommand{\ds}{\displaystyle}
\newtheorem{theorem}{Theorem}[section]
\newtheorem{thm}[theorem]{Theorem}

\newtheorem{lem}[theorem]{Lemma}
\newtheorem{cor}[theorem]{Corollary}

\def\smallR{\hbox{\scriptsize I\kern-.23em{R}}}
\def\0{\hbox{$\mit I$\kern-.70em$\mit O$}}
\def\r{I\kern-.277em R}

\usepackage{amsfonts}

\begin{document}

\title{Exponentially Accurate Semiclassical
Tunneling Wave Functions in One Dimension}

\author{
Vasile Gradinaru\\
Seminar for Applied Mathematics\\
ETH Z\"urich\\
CH--8092 Z\"urich, Switzerland,\\[15pt]
George A. Hagedorn\thanks{Partially
Supported by National Science Foundation
Grants DMS--0303586 and DMS--0600944.}\\
Department of Mathematics~ and\\
\hspace{-29pt}Center for Statistical Mechanics, Mathematical Physics,
and Theoretical Chemistry\\
Virginia Polytechnic Institute and State University\\
Blacksburg, Virginia~ 24061--0123,~ U.S.A.\\[15pt]
\and
Alain Joye\\
Institut Fourier\\ Unit\'e Mixte de Recherche CNRS--UJF 5582\\
Universit\'e de Grenoble I,\quad
BP 74\\
F--38402 Saint Martin d'H\`eres Cedex, France
}

\maketitle

\centerline{\it Dedicated to the memory of
our friend and colleague Pierre Duclos.}

\vskip 8mm
\begin{abstract}
We study the time behavior of wave functions involved in tunneling
through a smooth potential barrier in one dimension in the semiclassical
limit. We determine the leading order component of the wave function that
tunnels. It is exponentially small in $1/\hbar$. For a wide variety of
incoming wave packets, the leading order tunneling component is Gaussian
for sufficiently small $\hbar$. We prove this for both the large time
asymptotics and for moderately large values of the time variable.
\end{abstract}

\newpage
\baselineskip=20pt

\section{Introduction}
The goal of this paper is to study the semiclassical limit of
solutions to the one--dimensional time--dependent Schr\"odinger equation
that involve tunneling through a simple potential barrier.
Numerical simulations that illustrate our results
are presented in \cite{GHJjcp}.
A related wave packet ``spawning'' algorithm is also presented there.

%
%
Specifically, we consider solutions to
\be\label{tdsch}
i\,\hbar\,\frac{\partial\Psi}{\partial t}\ =\
\left(\,-\ \frac{\hbar^2}{2}\ \frac{\partial^2\phantom{i}}{\partial x^2}\
+\ V(x)\,\right)\,\Psi,
\qquad\quad
\Psi(\cdot,\,t,\,\hbar)\in L^2(\R),
\ee
for small values of $\hbar$, where the potential $V$ is an analytic
function that represents a barrier. Our goal is to present formulas
for the part of the wave function that has tunneled through the barrier.

%
%
\subsection{A Qualitative Synopsis of our Results}

Our results, stated in Theorems \ref{bark} and \ref{mod}
and Corollary \ref{growl}, are quite technical, so we begin with
an informal, qualitative discussion of a special case.

Suppose $V(x)$ is a very simple, analytic bump function that
tends sufficiently rapidly to zero as
$x$ tends to $+\infty$ and $-\infty$.
For a wave coming in from the left, we choose
generalized eigenfunctions that satisfy
$$
\psi(k,\,x,\,\hbar)\quad \approx\quad
\left\{\
\begin{array}{lcl}
e^{ikx/\hbar}\ +\ R(k,\,\hbar)\ e^{-ikx/\hbar}
&\qquad\mbox{as}\quad &x\to-\infty
\\[3mm]
T(k,\,\hbar)\ e^{ikx/\hbar}
&\qquad\mbox{as}\quad &x\to+\infty,
\end{array}\right.
$$
with $k>0$.

We take superpositions of these, with energies below the top of
the barrier $V$, to form wave packets and let them evolve.
For a wide class of such superpositions, we have the following:
\begin{enumerate}
\item[{\bf 1.}~]
If the average momentum of the incoming wave packet is~ $\eta$,~
then the transition probability for tunneling is strictly
greater than~ $|T(\eta,\,\hbar)|^2$.
\item[{\bf 2.}~]
The average momentum of the tunneling wave packet is
strictly greater than~ $\eta$.
\item[{\bf 3.}~]
The leading term for the tunneling wave packet for small~
$\hbar$~ is a complex Gaussian.
\end{enumerate}

Our main results underpin the numerical spawning algorithm
of \cite{GHJjcp} that describes semiclassical tunneling.
The qualitative item 3 is crucial to the algorithm, whose
quantitative information is determined numerically.

Intuitively, items 1, 2, and 3 are easy to understand.
When $\hbar$ is small, the function $|T(k,\,\hbar)|$ increases
extremely rapidly as $k$ increases.  For $k$ near $\eta$, it typically
grows like $C\ \exp(\alpha(k-\eta)/\hbar)$, with $\alpha>0$.
Thus, higher momentum components of the wave packet are much more likely
to tunnel than the average momentum components.
Items 1 and 2 are consequences of this observation.

One can understand item 3 and learn more about item 2 by examining the
transmitted wave packet in momentum space after tunneling has occurred.
For example, if the incoming wave packet in momentum space is chosen
to be asymptotic to one of the semiclassical wave packets~ $\phi_j$~
of \cite{raise},
$$
e^{-itk^2/(2\hbar)}\
2^{-j/2}\ (j!)^{-1/2}\ \pi^{-1/4}\ \hbar^{-1/4}\
H_j((k-\eta)/\hbar^{1/2})\
\exp(-(k-\eta)^2/(2\,\hbar)),
$$
then the transmitted wave packet behaves like
$$
C\ \exp(\alpha(k-\eta)/\hbar)\
e^{-itk^2/(2\hbar)}\
2^{-j/2}\ (j!)^{-1/2}\ \pi^{-1/4}\ \hbar^{-1/4}\
H_j((k-\eta)/\hbar^{1/2})\
\exp(-(k-\eta)^2/(2\,\hbar)).
$$
This equals
$$
e^{-itk^2/(2\hbar)}\
C\ e^{+\alpha^2/(2\hbar)}\
2^{-j/2}\ (j!)^{-1/2}\ \pi^{-1/4}\ \hbar^{-1/4}\
H_j((k-\eta)/\hbar^{1/2})\
\exp(-(k-\eta-\alpha)^2/(2\,\hbar)),
$$
The Gaussian factor is large only near $k=\eta+\alpha$, and near there,
the Hermite polynomial behaves like its leading term,
$2^j\ \alpha^j/\hbar^{j/2}$. This product is asymptotically another
Gaussian with momentum near~ $\eta+\alpha$~ for sufficiently small~ $\hbar$.

We note that the $C$ in these expressions behaves like~
$e^{-K/\hbar}$,~ so none of this can be determined by a perturbation
expansion in powers of~ $\hbar$.

\vskip 5mm
To obtain quantitative results, we must insert energy cut offs
and deal with many other technicalities,
but the description above gives an intuitive summary of our results.

The precise statements of our results are presented in Section 2.~
Theorem \ref{bark} presents the very large time behavior
of the tunneling wave function.
Theorem \ref{mod} and Corollary \ref{growl} describe the
behavior of the tunneling wave function for all times shortly
after the tunneling has occurred.


%
%
\vskip 5mm
\section{A More Precise Description of the Problem}
\setcounter{equation}{0}

We consider wave packets that have their energy  localized
in an interval~ $\Delta=[E_1,\,E_2]$,~ and we
assume the potential $V$ satisfies the following hypotheses:
\begin{enumerate}
\item[{\bf i)}]~ $x\mapsto V(x)$~ is real analytic in the strip~
$S_\alpha\,=\,\{z:\, |\mbox{Im}\, z |\,\leq\,\alpha\}$,~ for some~ $\alpha>0$.
\item[{\bf ii)}]~ There exist~ $V(\pm\infty)\in (-\infty,\,E_1)$,~
$\nu>1/2$,~ and~ $c<\infty$~
such that
\be\label{scatpot}
\limsup_{\mbox{\scriptsize Re}\,z\ra\pm\infty}\
\left(\,|\mbox{Re}\,z|^{2+\nu}\ \,
\sup_{\left|\mbox{\scriptsize Im}\,z\right|\leq \alpha}\,|V(z)-V(\pm\infty)|\,
\right)\ \,<\ \,c.
\ee
\item[{\bf iii)}]~ For any~ $E\in\Delta$,~ the function~ $V(x)-E$~
has exactly two simple zeros,~ $x_0(E)<x_1(E)$ with $x_1(E)<1$.
\end{enumerate}

Under these hypotheses, we can decompose our solutions as superpositions
of generalized eigenvectors of
$$
H(\hbar)\ =\ -\ \frac{\hbar^2}{2}\ \frac{\partial^2\phantom{i}}{\partial x^2}\
+\ V(x)
$$
whose energy lies in $\Delta$.
%

Since we are interested in the tunneling process, we assume that in the distant
past, the wave packet was a scattering state, coming in from the left of the
potential barrier. Our goal
is to describe the leading order behavior of the wave packet on the right of the
potential barrier for small $\hbar$ and sufficiently large positive times.
This tunneling wave is well-known to have exponentially small norm in $1/\hbar$.
We determine its leading order component and show that for a wide variety of
incoming states, it is a Gaussian.

Analogous results for exponentially small reflected waves when the energy
is strictly  above a potential bump are presented in \cite{bjt09}.
Similar results for non--adiabatic transitions in the Born--Oppenheimer
approximation are presented in \cite{hj05} and \cite{jm05}.

%
%
\subsection{Generalized Eigenfunction Expansions}
For any fixed energy~ $\ds E<\max_{x\in\R}\,V(x)$,~ we
let~ $\psi(x,\,E,\,\hbar)$~ be the solution to the stationary
Schr\"odinger equation
\be\label{ssch}
-\ {\hbar^2}\ \frac{\partial^2\psi}{\partial x^2}\ \,=\
\,2\ (E-V(x))\ \psi\ \,
\equiv\ \,p^2(x,E)\ \psi,
\ee
that we construct below.
Here,
$$
p(x,\,E)\ =\ \sqrt{2\,(E-V(x))\,}\ \,>\ \,0,
\qquad\mbox{for}\qquad |x|\gg 1,
$$
is the classical momentum at energy $E$.
The turning points~ $x_0(E)<x_1(E)$,~ given by the two
solutions of~ $p(x,\,E)=0$~ are branch points of~ $p(x,\,E)$,~
viewed as a multi-valued analytic function of $x$.
We use the multivaluedness of this function in the analysis below.

We consider wave functions
\be\label{superp}
\Psi(x,\,t,\,\hbar)\ \,=\ \,\int_{\Delta}\ Q(E,\,\hbar)\
e^{-\,i\,t\,E/\hbar}\ \psi(x,\,E,\,\hbar)\ dE,
\ee
for some sufficiently regular energy density $Q(E,\,\hbar)$ defined
on $\Delta$. Such a function is a solution of (\ref{tdsch})
under the hypotheses we impose below on the energy density $Q$.\\

We shall derive a space--time description of the exponentially
tunneling wave to leading order as~ $\hbar\ra 0$~ for large positive times,
when this wave is far enough from the potential bump.\\

The energy densities~ $Q(E,\,\hbar)$~ we choose are
sharply peaked at a specific value\\
$E_0\in (E_1,\,E_2)$.~
Specifically, we consider
\be\label{defq}
Q(E,\,\hbar)\ \,=\ \,e^{-\,G(E)/\hbar}\
e^{-\,i\,J(E)/\hbar}\ P(E,\,\hbar),
\ee
where:
\begin{enumerate}
\item[{\bf (C1)}~] The real-valued function $G\ge 0$ is in
$C^3(\Delta)$, is independent of $\hbar$, and has a unique
non-degenerate minimum value of~ $0$~ at~ $E_0\in(E_1,\,E_2)$.~
This implies that
$$
G(E)\ =\ g\,(E-E_0)^2/2\ +\ O\left((E-E_0)^3\right),
\quad\mbox{where}\quad g>0.
$$
\item[{\bf (C2)}~] The real-valued function~ $J$~ is in $C^3(\Delta)$.
\item[{\bf (C3)}~] The complex-valued function~ $P(E,\,\hbar)$~ is in
$C^1(\Delta)$ and satisfies
$$
\sup_{E\in\Delta\atop \varepsilon \geq 0}\
\left|\,\frac{\partial^n}{\partial E^n}\,P(E,\,\hbar)\,\right|\ \,
\leq\ \,C_n,\qquad\mbox{for}\qquad n=0,\,1.
$$
\end{enumerate}

%
%
\subsection{The Specific Generalized Eigenfunction~ $\psi(x,\,E,\,\hbar)$}

We first recast the eigenvalue problem for $H(\hbar)$
as a first order system of linear equations.
We then use the analyticity of
the potential and consider the  extensions of these equations
to the complex $x$-plane to perform our asymptotic analysis.
The function $p(\cdot,\,E)$ has branch points at
$x_0(E)$ and $x_1(E)$ which we have to select in a consistent way.
Note the turning points~ $x_0(E)<x_1(E)$~ are separated for all~
$E\in\Delta$.

We initially confine $x$ to the real axis and define
\be\label{pr}
p_R(x,\,E)\ \,=\ \,\sqrt{2\,|E-V(x)|\,}\ \,\ge\ \,0.
\ee
We begin our analysis for~ $x>x_1(E)$,~ where~ $p_R(x,\,E)$~ is
the classical momentum.

Suppose~ $\zeta$~ satisfies the ordinary differential equation
(\ref{ssch}).~
For $x>x_1(E)$, we define
\be\label{clos}
\Phi(x)\ =\ \left(\begin{array}{c}
\zeta(x)\\[2mm] i\,\hbar\,\zeta'(x)
\end{array}\right),
\ee
where $\phantom{I}'$ denotes the derivative with respect to~ $x$.~
We note that~ $\Phi$~ satisfies
$$
i\,\hbar\,\Phi'(x)\ =\ \pmatrix{0 & 1 \cr p^2(x,\,E) & 0}\ \Phi(x)\
\equiv\ A(x,\,E)\ \Phi(x).
$$
In order to apply the WKB method, we expand~
$\Phi(x)$~ in terms of the instantaneous eigenvectors of the generator of~
$A(x,\,E)$.~ Specifically, we choose
\be\nonumber\label{eigvec}
\ffi_1(x)\ =\ \left(\begin{array}{c}
\frac 1{\sqrt{p_R(x,\,E)}}\\[4mm] \sqrt{p_R(x,\,E)}
\end{array}\right)
\quad\mbox{and}\quad
\ffi_2(x)\ =\ \left(\begin{array}{c}
\frac1{\sqrt{p_R(x,\,E)}}\\[4mm] -\,\sqrt{p_R(x,\,E)}
\end{array}\right)
\ee
to be the eigenvectors associated with the eigenvalues
$p_R(x,\,E)$ and $-p_R(x,\,E)$, respectively.

We decompose~ $\Phi(x)$~ as
\be\label{phidec}
\Phi(x)\ =\ c_1(x)\ e^{-i\int_{x_1(E)}^xp_R(y,\,E)\,dy/\hbar}\
\ffi_1(x)\ +\
c_2(x)\ e^{i\int_{x_1(E)}^xp_R(y,\,E)\,dy/\hbar}\ \ffi_2(x),
\ee
where~ $c_1(x)$~ and~ $c_2(x)$~ are complex-valued
coefficients that satisfy
\bea\label{eqco}
\hspace{-1cm}
\left(\begin{array}{c}
c_1'(x)\\[2mm] c_2'(x)
\end{array}\right)
\ =\ \frac{p'_R(x,\,E)}{2\,p_R(x,\,E)}\,
\left(\begin{array}{cc}
0&e^{2i\int_{x_1(E)}^xp_R(y,\,E)\,dy/\hbar}\\[3mm]
e^{-2i\int_{x_1(E)}^xp_R(y,\,E)\,dy/\hbar}&0
\end{array}\right)\,
\left(\begin{array}{c}
c_1(x)\\[2mm] c_2(x)
\end{array}\right)
\eea
These coefficients depend on~ $E$,~ and on sets where~ $p(x,\,E)$~
is real-valued, they satisfy \cite{jp95}
\begin{enumerate}
\item[$\bullet$~~]
$|c_1(x)|^2-|c_2(x)|^2$\quad is independent of $x$,\qquad and
\item[$\bullet$~~]
$\ds \left(\begin{array}{c} c_1(x)\\[1mm] c_2(x)
\end{array}\right)$\quad
is a solution to (\ref{eqco})\quad
if and only if\quad
$\left(\begin{array}{c}
\overline{c_1(x)}\\[1mm]\overline{c_2(x)}
\end{array}\right)$\quad is a solution to (\ref{eqco}).
\end{enumerate}


This decomposition  allows us to write the
solution to (\ref{ssch}) for large positive $x$ as
\bea\nonumber
\zeta(x,\,E,\,\hbar)&=&
c_1(x,\,E,\,\hbar)\ e^{-i\int_{x_1(E)}^xp_R(y,\,E)\,dy/\hbar}\
\ffi_1^{(1)}(x,\,E,\,\hbar)
\\[4mm]\label{decomp}
&&
+\quad c_2(x,\,E,\,\hbar)\ e^{i\int_{x_1(E)}^xp_R(y,\,E)\,dy/\hbar}\
\ffi_2^{(1)}(x,\,E,\,\hbar),
\eea
where the~ $E$~ and~ $\hbar$~ dependence is explicit, and~
$\ffi_j^{(1)}$~ denotes the first component of~ $\ffi_j$.

When $x$ is smaller than $x_0(E)$, so that the classical momentum again
equals $p_R(x,E)$, a similar expansion is valid. However, we take a different
phase convention when $x<x_0(E)$:
\bea\nonumber\
\zeta(x,\,E,\,\hbar)&=&
d_1(x,\,E,\,\hbar)\ e^{-i\int_{x_0(E)}^xp_R(y,\,E)\,dy/\hbar}\
\ffi_1^{(1)}(x,\,E,\,\hbar)
\\[4mm]\label{dcoef}
&&+\quad
d_2(x,\,E,\,\hbar)\ e^{i\int_{x_0(E)}^xp_R(y,\,E)\,dy/\hbar}\
\ffi_2^{(1)}(x,\,E,\,\hbar).
\eea
where the coefficients~ $d_1$~ and~ $d_2$~ satisfy a differential
equation similar to (\ref{eqco}).

We need
to connect the coefficients
$c_1$ and $c_2$ to the coefficients $d_1$ and $d_2$.
One commonly used technique consists of solving a similar equation
for $x_0(E)<x<x_1(E)$
and matching the solutions to those we just described.
Instead we use the complex WKB method which allows us to work with just
one equation, but requires an analytic framework.
See \cite{BM} for various possible approaches.\\

We only want to have an outgoing wave on the right
and are not currently worrying about normalization,
so we consider the asymptotic conditions
\be\label{incond}
c_1(+\infty,\,E,\,\hbar)\ =\ 0,\qquad\mbox{and}\qquad
c_2(+\infty,\,E,\,\hbar)\ =\ 1.
\ee
Then as in \cite{hj05} and \cite{jm05}, we have
\bea\nonumber
p_R(\pm\infty,\,E)&>&0,
\\[4mm]\nonumber
\ffi_j(\pm\infty,\,E)&=&
\left(\begin{array}{c}
\frac 1{\sqrt{p_R(\pm\infty,\,E)}}
\\[5mm]
(-1)^{j+1}\,\sqrt{p_R(\pm\infty,\,E)}
\end{array}\right),
\\[4mm]\nonumber\hspace{-4pt}
\int_{x_1(E)}^x\,p_R(y,\,E)\,dy&=&
(x-x_1(E))\ p_R(\infty,\,E)
\ +\ \int_{x_1(E)}^{\infty}\,(p_R(y,\,E)-p_R(\infty,\,E))\,dy
\\[4mm]\nonumber
&&+\quad
O\left(|x|^{-1-\nu}\right),
\qquad\mbox{as}\quad x\to\infty,\qquad\quad\mbox{and}
\\[4mm]\nonumber\hspace{-4pt}
\int_{x_0(E)}^x\,p_R(y,\,E)\,dy&=&
(x-{x_0(E)})\ p_R(-\infty,\,E)\ +\
\int_{x_0(E)}^{-\infty}\,(p_R(y,\,E)-p_R(-\infty,\,E))\,dy
\\[4mm]\nonumber
&&+\quad
O\left(|x|^{-1-\nu}\right)
\qquad\mbox{as}\quad x\to -\infty.
\eea

The error estimates here are uniform for~ $E\in\Delta$.

We thus have an incoming wave asymptotically described
for large negative~ $x$~ by
$$
d_2(-\infty,\,E,\,\hbar)\ \,
\frac{e^{i\int_{x_0(E)}^{-\infty}(p_R(y,\,E)-p_R(-\infty,\,E))\,dy/\hbar}}
{\sqrt{p_R(-\infty,\,E)\,}}\ \,
e^{-ip_R(-\infty,\,E)x_0(E)/\hbar}\ \,
e^{i\,p_R(-\infty,\,E)\,x/\hbar}.
$$
The reflected wave is asymptotically described for large negative~ $x$~ by
$$
d_1(-\infty,\,E,\,\hbar)\ \,
\frac{e^{-i\int_{x_0(E)}^{-\infty}(p_R(y,\,E)-p_R(-\infty,\,E))\,dy/\hbar}}
{\sqrt{p_R(-\infty,\,E)\,}}\ \,
e^{+ip_R(-\infty,\,E)x_0(E)/\hbar}\ \,
e^{-\,i\,p_R(-\infty,\,E)\,x/\hbar}.
$$
The transmitted wave is asymptotically described for large positive~ $x$~ by
\be\label{unret}
\frac{e^{i\int_{x_1(E)}^{+\infty}(p_R(y,\,E)-p_R(+\infty,\,E))\,dy/\hbar}}
{\sqrt{p_R(\infty,\,E)\,}}~\,
e^{-ip_R(-\infty,\,E)x_1(E)/\hbar}\ \,
e^{i\,p_R(+\infty,\,E)\,x/\hbar}.
\ee

We obtain the solution~ $\psi(x,\,E,\,\hbar)$~ that we use in (\ref{superp}) by
normalizing the incoming flux. To do so, we simply divide
the whole solution~ $\zeta(x,\,E,\,\hbar)$~ by the constant
$$
d_2(-\infty,\,E,\,\hbar)~\,
\frac{e^{i\int_{x_0(E)}^{-\infty}(p_R(y,\,E)-p_R(-\infty,\,E))\,dy/\hbar}}
{\sqrt{p_R(-\infty,\,E)\,}}~\,
e^{-ip_R(-\infty,\,E)x_0(E)/\hbar}.
$$

We obtain our main results by analyzing the tunneling wave by
studying the large $x$ and $t$ asymptotics
of (\ref{superp}) with this~ $\psi(x,\,E,\,\hbar)$~ and an any appropriate
choice of~ $Q(E,\,\hbar)$.



%
%
\subsection{Complex WKB analysis}

We need to compute the asymptotic behavior, as $x\ra-\infty$,
of the solution to (\ref{eqco}) that satisfies (\ref{incond}).
We do this by applying the complex WKB method in order to
avoid matching of the solutions at the real turning points
$x_0(E)$ and $x_1(E)$, where the equation is ill-defined.
So, we consider (\ref{ssch}) and (\ref{eqco}) in the strip $S_\alpha$
in the complex plane containing the real axis, with possible branch cuts
at $x_0(E)$ and  $x_1(E)$.
We now replace the variable $x$ by $z$, to emphasize that the variable
is no longer restricted to the real line.
The solution to (\ref{ssch}) is analytic for $z\in S_\alpha$, but
the solution to (\ref{eqco}) has singularities at the turning points.
As~ $\mbox{Re}\,z$~ tends to~ $+\infty$~ in $S_\alpha$,
our assumptions on the behavior of the potential
ensure that the asymptotic values of the coefficients $c_j(z)$
are independent of~ $\mbox{Im}\,z$.~
We can thus start the integration of (\ref{eqco}) above the real axis
at the extreme right of the strip, with asymptotic boundary
data (\ref{incond}).
Also, our assumptions imply the existence of two $C^1$ paths from the right
end of the strip  $S_\alpha$
to its left end, with one of them, $\gamma_a$,
passing above the two turning points,
and the other, $\gamma_b$, passing between them.
We parameterize these with $t\in\R$ and assume they satisfy
$$
\gamma_\#(t)\in S_\alpha,\qquad\mbox{with}\qquad
\lim_{t\ra\pm\infty}\ \mbox{Re}\ \gamma_\#(t)\ =\ \mp\infty,\qquad
\mbox{and}\qquad\sup_{t\ra\pm\infty}\ |\dot \gamma_\#(t)|\ <\ \infty,
$$
where $\#$ stands for $a$ or $b$.
Note that~ $\mbox{Im}\,\gamma_a(t) >0$~ for all $t$, whereas~
$\mbox{Im}\,\gamma_b$~ changes sign exactly once and
is positive for $t$ large and negative.
Finally, and this is the main property of these paths,
the imaginary part of~ $\int_{x_1(E)}^z\,p(z',\,E)\,dz'$~
along~ $\gamma_\#$~ is decreasing.
Such paths are called dissipative.

The existence of these dissipative paths is proved as in
\cite{jkp} and \cite{j97}.
Close enough to the real axis, there exist level
lines of the function~
$\mbox{Im}\,\int_{x_1(E)}^z\,p(z',\,E)\,dz'$~
which are essentially parallel
to the real axis for~ $\mbox{Re}\,z\geq x_1(E)$~ and~
$\mbox{Re}\,z\leq x_0(E)$.~
For~
$x_0(E)\leq \mbox{Re}\,z \leq x_1(E)$,\\
these lines can be connected by means of level lines of~
$\mbox{Re}\,\int_{x_1(E)}^z\,p(z',\,E)\,dz'$,~ along which\\
$\mbox{Im}\,\int_{x_1(E)}^z\,p(z',\,E)\,dz'$~ is strictly decreasing.
As a local analysis reveals, the connections can be made in a
$C^1$ fashion without losing the dissipativity property.
%
It is readily seen by inspection, that $\gamma_a$
can be constructed this way. For $\gamma_b$,
one starts as for $\gamma_a$, and between
$x_0(E)$ and $x_1(E)$, one uses a level line of~
$\mbox{Im}\, \int_{x_1(E)}^z\, p(z',\,E)\,dz'$~ to cross the real
axis. Then one connects to a level line of~
$\mbox{Re}\, \int_{x_1(E)}^z\, p(z',\,E)\,dz'$~ and proceeds as
described above to connect to $-\infty$ below the real axis.

\bigskip
We can integrate (\ref{eqco}) along these two different paths
and compare the solutions for large negative~ $\mbox{Re}\,z$~ with
$z\in S_\alpha$.
These two integrations
describe the same solution to (\ref{ssch}) since it is analytic.
Moreover, the asymptotic values of the coefficients as $\hbar\ra 0$
can be controlled, because both these paths are dissipative.

We choose two specific branches, $p^a(z,\,E)$ and $p^b(z,\,E)$,
of the multivalued function $p(z,\,E)$.
For $p^a(z,\,E)$ we place vertical branch cuts below the real axis,
extending down from $x_0(E)$ and $x_1(E)$. For $p^b(z,\,E)$ we place
a vertical branch cut below the real axis extending down from $x_1(E)$,
and a vertical branch cut above the real axis extending up from $x_0(E)$.
These two functions are then uniquely determined in their respective
regions $S_\alpha^a$ and $S_\alpha^b$ by the requirement that they both
equal $p_R(z,\,E)$ when $z$ is real and greater than $x_1(E)$.

The following are satisfied when $z$ is on the real axis:
%
$$
p^a(x,\,E)\quad=\quad\left\{
\begin{array}{ccl}
p_R(x,\,E)\ >\ 0,&\mbox{~if~}& x>x_1(E)
\\[4mm]
e^{i\pi/2}\ p_R(x,\,E),&\mbox{~if~}& x_0(E)<x<x_1(E)
\\[4mm]
-\ p_R(x,\,E)\ <\ 0,&\mbox{~if~}& x<x_0(E).
\end{array}\right.
$$
and
%
$$
p^b(x,\,E)\quad=\quad\left\{
\begin{array}{ccl}
p_R(x,\,E)\ >\ 0,&\mbox{~if~}& x>x_1(E)
\\[4mm]
e^{i\pi/2}\ p_R(x,\,E),&\mbox{~if~}& x_0(E)<x<x_1(E)
\\[4mm]
\phantom{-\ }p_R(x,\,E)\ >\ 0,&\mbox{~if~}& x<x_0(E).
\end{array}\right.
$$
For~ $\#=a,\,b$,~ the function~ $p^\#(z,\,E)$~ is analytic
in a neighborhood of the path~ $\gamma_\#$.

We define~
%
$\ffi_j^\#(z,\,E)$,~ to be the analytic continuation in $S_\alpha^\#$
of the vector~ $\ffi_j(z,\,E)$~ defined in (\ref{eigvec}).
We also define~ $\ds\int_{x_1(E)}^z\,p^\#(y,\,E)\,dy$~
to be the analytic continuation in  $S_\alpha^\#$ of
$\ds\int_{x_1(E)}^z\,p(y,\,E)\,dy$,~ which is already specified for $x>x_1(E)$.

%
%
\begin{lem}\label{21}\quad
For any real~ $z<x_0(E)$,~ the following are satisfied:
\bea\nonumber
&&i\ \ffi_1^a(z)\ =\ \ffi_2^b(z)\ =\
\left(\,\begin{array}{c}
\frac{1}{\sqrt{p_R(z,\,E)\,}}\\[4mm]  -\ \sqrt{p_R(z,\,E)\,}
\end{array}\right),
\\[4mm]\nonumber
&&i\ \ffi_2^a(z)\ =\ \ffi_1^b(z)\ =\
\left(\,\begin{array}{c}
\frac{1}{\sqrt{p_R(z,E)\,}}\\[4mm]  \sqrt{p_R(z,\,E)\,}
\end{array}\right),
\\[4mm]\label{phase}
&&\int_{x_1(E)}^z\,p^b(y,\,E)\,dy\ =\
i\ \int_{x_1(E)}^{x_0(E)}\,p_R(y,\,E)\,dy\ +\
\int_{x_0(E)}^z\,p_R(y,\,E)\,dy
\\[4mm]\nonumber
&&\int_{x_1(E)}^z\,p^a(y,\,E)\,dy\ +\
\int_{x_1(E)}^z\,p^b(y,\,E)\,dy\ =\
-\,2\,i\ \int_{x_0(E)}^{x_1(E)}\,p_R(y,\,E)\,dy.
\eea
\end{lem}
\vskip 5mm
{\bf Proof}\quad
We simply follow the analytic continuations of~ $p$~
in the respective domains.\hfill\ep\\

\noindent{\bf Remarks}
\begin{enumerate}
\item[{\bf i)}~]
The quantity~ $\ds 2\,\int_{x_0(E)}^{x_1(E)}\,p_R(y,\,E)\,dy$~
can be expressed as a contour integral
$$
K(E)\ =\ 2\,\int_{x_0(E)}^{x_1(E)}\,p_R(y,\,E)\,dy\ =\
\int_\gamma\,p(z,\,E)\,dz\ >\ 0,
$$
where $\gamma$ is a simple negatively oriented loop
around the two turning points, and~
$p(z,\,E)$ is the analytic continuation of~ $p_R(x,\,E)$~
for~ $x>x_0(E)$.~ This shows that~ $K(E)$~ is analytic for~ $E$~
in a complex neighbourhood of the energy window~ $\Delta$.
\item[{\bf ii)}~]
Equation (\ref{phase}) shows that the analytic continuations~
$(c_1^b(z),\,c_2^b(z))$~ of the coefficients coincide
with~ $(d_1(z),\,d_2(z))$~ in (\ref{dcoef}) for~ $z<x_0(E)$,~
up to multiplicative constants.
\end{enumerate}

\vskip 5mm
Coming back to the differential equation (\ref{eqco}),
we denote the analytic continuations of its solutions
in $S_\alpha^\#$ as $(c_1^\#, c_2^\#)$.
We consider the analytic function $\Phi$ for $z<x_0(E)$
and the two different analytic
continuations of its decomposition  (\ref{phidec}) at $z$.
These two representations must agree. This and Lemma \ref{21}
imply the following:

%
%
\begin{lem}\quad For any~ $z<x_0(E)$,~ we have
\bea\nonumber
i\ c_2^b(z,\,E,\,\hbar)\ e^{K(E)/\hbar}
&=&c_1^a(z,\,E,\,\hbar)
\\[4mm]\nonumber
i\ c_1^b(z,\,E,\,\hbar)\ e^{-K(E)/\hbar}
&=&c_2^a(z,\,E,\,\hbar).
\eea
\end{lem}

\vskip 5mm\noindent
{\bf Remark}\quad
The identities in the two previous lemmas are actually true
for any $z$ with\\ $\mbox{Re}\,z<x_0(E)$.

\vskip 5mm
The WKB analysis of (\ref{eqco}) along the dissipative paths
$\gamma^\#$ and assumption (\ref{scatpot}) now yield the following lemma,
as shown in \cite{jkp}, \cite{j97}, \cite{hj05}, \cite{jm05}.

%
%
\begin{lem}\label{23}\quad We have
\bea\nonumber
c_2^{a}(-\infty,\,E,\,\hbar)
&=&1\ +\ O_{E}(\hbar),\qquad\mbox{as}\quad\hbar\ra 0,
\\[3mm]\nonumber
c_2^{b}(-\infty,\,E,\,\hbar)
&=&1\ +\ O_{E}(\hbar),\qquad\mbox{as}\quad\hbar\ra 0,\qquad\mbox{and}
\\[3mm]\nonumber
 c_j^ \#(x,\,E,\,\hbar)
&=&c_j^ \#(\pm\infty,\,E,\,\hbar)\ +\ O_{E}(1/|x|^{1+\nu})\qquad
\mbox{as}\quad x\ra\pm \infty,
\eea
where the remainder terms are analytic in~ $E$,~ for~ $E$~
in a complex neighborhood of the real set~ $\Delta$.~
Moreover,~
$\ds\frac{d\phantom{i}}{dE}\ c_2^{b}(-\infty,\,E,\,\hbar)$~
and the ~ $O_{E}(1/|x|^{\nu})$~ are uniformly bounded as~ $\hbar\ra 0$.
\end{lem}

\vskip 5mm
As a consequence of this lemma, for~ $x \gg 1$,~ we have
\bea\nonumber
\zeta(x,\,E,\,\hbar)&=&
\frac{e^{i(\int_{x_1(E)}^\infty\,(p_R(y,\,E)-p_R(\infty,\,E)\,dy)/\hbar}\
e^{-ix_1(E)p_R(\infty,\,E)/\hbar}}{\sqrt{p_R(\infty,\,E)\,}}\quad
e^{i\,p_R(+\infty,\,E)\,x/\hbar}
\\[4mm]\nonumber
&&\times\quad
\left(\,1+\,O_{E}(\hbar)\ +\ O_E(1/(\hbar\,|x|^{1+\nu})\,\right)
\eea
and, for~ $x \ll -1$,~ we have
\bea\nonumber\hspace{-1cm}
\zeta(x,\,E,\,\hbar)&=&
-\ i\ e^{K(E)/(2\hbar)}\ \,
\frac{e^{i(\int_{-\infty}^{x_0(E)}(p_R(s,E)-p_R(-\infty,E)ds)/\hbar}\
e^{ix_0(E)p_R(-\infty,E)/\hbar}}
{\sqrt{p_R(-\infty,\,E)\,}}\ \,e^{-\,i\,p_R(-\infty,\,E)\,x/\hbar}
\\[4mm]\nonumber
&&+\quad e^{K(E)/(2\hbar)}\ \,
\frac{e^{-i(\int_{-\infty}^{x_0(E)}(p_R(s,E)-p_R(-\infty,E)ds)/\hbar}\
e^{-ix_0(E)p_R(-\infty,E)/\hbar}}
{\sqrt{p_R(-\infty,\,E)\,}}\ \,e^{i\,p_R(-\infty,\,E)\,x/\hbar}
\\[4mm] \nonumber
&&+\quad e^{K(E)/(2\hbar)}\
\left(\ e^{i\,p_R(-\infty,\,E)\,x/\hbar}\ +\
e^{-\,i\,p_R(-\infty,E)\,x/\hbar}\ \right)
\\[4mm]\label{wkb}
&&\qquad\times\quad
\left(\,O_{E}(\hbar)\ +\ O_E(1/(\hbar\,|x|^{1+\nu})\right).
\eea

\vskip 5mm
%
%
\subsection{Large Time Asymptotics of the Tunneling Wave Function}

\vskip 3mm
We now consider the large time behavior of the transmitted wave
packet.
We denote this wave packet by~ $\chi(x,\,t,\,\hbar)$.~
We construct it as a time-dependent superposition of the
normalized generalized wave functions~ $\psi(x,\,E,\,\hbar)$,~
where~ $\ds x>1>\max_{E\in\Delta}\,x_1(E)$.\\
%
The specific superposition we use is
\bea\label{defchi}
\chi(x,\,t,\,\hbar)&=&
\int_{\Delta}\ Q(E,\,\hbar)\
e^{-\,i\,t\,E/\hbar}\ \psi(x,\,E,\,\hbar)\ dE,
\eea
where for~ $\ds x>\max_{E\in\Delta}\,x_1(E)$,
\bea
\label{defchi2}
&&\psi(x,\,E,\,\hbar)
\\[3mm]\nonumber
&=&\frac{e^{-K(E)/(2\hbar)}\ \,\sqrt{p(-\infty,\,E)\,}\ \,
c_2^a(x,\,E,\,\hbar)\ \,e^{i\int_{x_1(E)}^x\,p_R(y,\,E)\,dy/\hbar}}
{\sqrt{p(x,\,E)\,}\ c_2^b(-\infty,\,E,\,\hbar)\
e^{-i\int_{-\infty}^{x_0(E)}(p_R(y,E)-p_R(-\infty,E))/\hbar}\
e^{-\,i\,p_R(-\infty,\,E)\,x_0(E)}}\,.
\eea
See Remark ii) after Lemma \ref{21}.
%
%
%
The asymptotics we have established show that
for large positive~ $x$,
\bea\nonumber
\chi(x,\,t,\,\hbar)&=&
\int_{\Delta}\ Q(E,\,\hbar)\ \,
\sqrt{\frac{p(-\infty,\,E)}{p(+\infty,\,E)}\,}\ \,
e^{-K(E)/(2\hbar)}\ \,
e^{i\,(p_R(\infty,\,E)\,x\,-\,E\,t)/\hbar}
\\[4mm]\nonumber
&&\hspace{58mm}\times\quad
e^{-\,i\,\omega(E)/\hbar}
\quad\left(\,1\,+\,r(x,\,E,\,\hbar)\,\right)\ \,dE,
\eea
where
\bea\nonumber
\omega(E)&=&
-\ \int_{x_1(E)}^\infty\,(p_R(y,\,E)-p_R(\infty,\,E))\,dy\
-\ \int_{-\infty}^{x_0(E)}\,(p_R(y,\,E)-p_R(-\infty,\,E))\,dy
\\[4mm]\nonumber
&&+\quad
p_R(-\infty,\,E)\,(x_1(E)-x_0(E)).
\eea
The error term~ $r(x,\,E,\,\hbar)$~ in this expression satisfies
$$
r(x,\,E,\,\hbar)\quad =\quad
O\left(\,\hbar\ +\ \frac{1}{\hbar\,|x|^{1+\nu}}\ +\
\frac{1}{|x|^{2+\nu}}\,\right)\quad =\quad
O\left(\,\hbar\ +\ \frac{1}{\hbar\,|x|^{1+\nu}}\,\right),
$$
uniformly for~ $E\in\Delta$, $x>1$ and $\hbar$ small enough.

We prove below that~ $\chi(x,\,t,\,\hbar)$~ asymptotically
propagates freely to the right for large positive~ $t$.

\bigskip
For~ $E\in \Delta$,~ we define
\be\label{gabe}
\alpha(E)\ =\ G(E)\ +\ K(E)/2\qquad\mbox{and}\qquad
\kappa(E)\ =\ J(E)\ +\ \omega(E),
\ee
where~ $G$~ and~ $J$~ are the functions in (\ref{defq}).~
We then have
\bea\label{woof}
\hspace{-1cm}&&\hspace{-7mm}\chi(x,\,t,\,\hbar)
\\[4mm]\nonumber
\hspace{-1cm}&=&
\int_\Delta\,P(E,\hbar)\
\sqrt{\frac{p(-\infty,\,E)}{p(+\infty,\,E)}\,}\
e^{-\alpha(E)/\hbar}\ e^{-\,i\,\kappa(E)/\hbar}\
e^{i\,(p_R(\infty,\,E)\,x\,-\,E\,t)/\hbar}\
(1\,+\,r(x,\,E,\,\hbar))\ dE.
\eea
We obtain the small $\hbar$ asymptotics of this integral
by Laplace's method. We first state a result concerning
the large $x$ and large $t$ behavior of $\chi(x,\,t,\, \hbar)$,
whose proof follows from the methods of \cite{hj05} and \cite{jm05},
but is easier. The detailed analysis of the $x$ and $t$ dependence
yields the following result.
See \cite{hj05} and \cite{jm05} for details.

%
%
\begin{thm}\label{bark}\quad
Assume the function~ $\alpha(E)$~ has a unique non-degenerate minimum
at\\ $E=E^*$~ in~ $\Delta$.~
Define~ $k(E)=p_R(\infty,\,E)$~ and~ $k^*=k(E^*)$.\\
There exist~ $\delta>0$~and~ $T_\hbar>0$,~
such that for~ $t>T_\hbar$ and all~$x>1$,
the transmitted wave satisfies
$$
\chi(x,\,t,\,\hbar)\quad=\quad\chi_{Gauss}^\infty(x,\,t,\,\hbar)\ +\
O\left(\,e^{-\,\alpha(E^*)/\hbar}\,\hbar^{3/4+\delta}\,\right),
$$
where the error term is measured in the $L^2$ norm,
uniformly for~ $t>T_\hbar$,~ and
\bea\nonumber
\chi_{Gauss}^\infty(x,t,\hbar)&=&
\sqrt{2\,\pi\,\hbar\,k^*\,}\ \,P(E^*,\,\hbar)\ \,
\sqrt{\frac{p(-\infty,\,E^*)}{p(+\infty,\,E^*)}\,}\ \,
e^{-\,\alpha(E^*)/\hbar}
\\[4mm]\nonumber
&&\times\quad
\frac{\exp\left\{\,-\,i\,(t\,E^*\,+\,\kappa(E^*)\,-\,k^*\,x)/\hbar\,\right\}}
{\left(\,\frac{d^2\phantom{i}}{dk^2}\,\alpha(E(k))|_{k^*}\,+\,
i\,(\,t\,+\,\frac{d^2\phantom{i}}{dk^2}\,\kappa(E(k))|_{k^*})\,\right)^{1/2}}
\\[4mm]\nonumber
&&\times\quad
\exp\,\left\{\,-\ \frac{(x\,-\,k^*(t+\kappa'(E^*)))^2}
{2\,\hbar\,\left(\,\frac{d^2\phantom{i}}{dk^2}\,\alpha(E(k))|_{k^*}\,+\,
i\,(\,t\,+\,\frac{d^2\phantom{i}}{dk^2}\,\kappa(E(k))|_{k^*})\,\right)}\,\right\}.
\eea
\end{thm}

\vskip 5mm\noindent
{\bf Proof Outline}\quad
The proof of this theorem is very technical, but is very similar
to ones presented for Theorem 5.1 of \cite{hj05} and Theorem 6 of \cite{jm05}.
One computes the leading term\\
$\chi_{Gauss}^\infty(x,\,t,\,\hbar)$~
by a rigorous version of Laplace's method, paying attention to the dependence
of the remainder terms on the parameters $x$ and $t$.
The $L^2$ norm of this leading term is
\be\label{l2n}\hspace{-7mm}
\hbar^{3/4}\ \pi^{3/4}\ \sqrt{2\,k^*\,}\ \,
e^{-\,\alpha(E^*)/\hbar}\ \,P(E^*,\hbar)\ \,
\sqrt{\frac{p(-\infty,\,E^*)}{p(+\infty,\,E^*)}\,}\ \,
\left(\,\left.
\frac{d^2\phantom{i}}{dk^2}\right|_{k=k^*}\alpha(E(k))\,\right)^{-1/4},
\ee
which is~ $O\left(\,\hbar^{3/4}\,e^{-\,\alpha(E^*)/\hbar}\,\right)$.~
By the methods of used in  \cite{hj05} and  \cite{jm05}, the $L^2$ norm
of the error term induced by~ $r(x,\,E,\,\hbar)$~
under the integral sign in (\ref{woof}) is of order~
$\hbar^{3/4+\delta}\,e^{-\alpha(E^*)}$,~ for some~ $\delta>0$,
provided $t$ is large enough.
Note that Lemma 1 and Proposition 5 of \cite{bjt09} allow us to get
better control of~ $T_\hbar$.~ (See below.)
\hfill\ep

\vskip 8mm
%
%
\subsection{The Transmitted Wave Function Shortly After Tunneling}

\vskip 4mm
Mimicking \cite{bjt09}, we shall now address the behavior
of the transmitted wave for finite values of $x$ and $t$,
shortly after the transmitted wave has left the region
where it emerges from the potential barrier.
Because the exponential decay of transmitted wave computed from the behavior
of $p(z,E)$ on the real
axis appears as a factor, see (\ref{defchi2}),
the analysis of the semicassical behavior of the coefficient
$c^a(x,\,E,\, \hbar)$ for finite values of $x>x_1(E)$
is simpler than in \cite{bjt09}.
By contrast, the appearance for finite $x$'s of the corresponding
exponentially small factor for the above barrier reflection required
to pass to the superadiabatic representation in \cite{bjt09}.
This is not necessary here so that we can stick to  the adiabatic basis
(\ref{decomp}).
We shall not, however, examine the more delicate details
of how the transmitted wave actually emerges from the barrier.
One should be able to address this much more technical topic by
adapting the results of \cite{bjt09}.

Let
$$
\rho(E)\ =\ -\ \int_{-\infty}^{x_0(E)}\,
(p_R(y,\,E)\,-\,p_R(-\infty,\,E))\,dy\quad-\quad
p_R(-\infty,\,E)\ x_0(E),
$$
$$
S(x,\,t,\,E)\ =\ -\,\int_{x_1(E)}^{x}\, p_R(y,E)\,dy\quad +\quad
\rho(E)\ +\ J(E)\ +\ E\,t,
$$
and
$$
P_0(x,\,E)\ =\ \frac{P(E,\,\hbar)\ \sqrt{p_R(-\infty,\,E)}\,}
{\sqrt{p_R(x,\,E)\,}}.
$$

In the region of moderately large positive $x$, but far from the potential
barrier, the transmitted wave is described by the following theorem,
which requires faster decay of the potential to its asymptotic value.

%
%
\begin{thm}\label{mod}\quad Let $\nu >21/2$.~
There exist~ $\delta>0$,~ $\tau>0$,~ $C>0$,~ and~ $\beta>0$,~
such that for all~ $t>\tau$~ and sufficiently small~ $\hbar$,
$$
\chi(x,\,t,\,\hbar)\ =\
\left\{\ \begin{array}{cl}
\chi_{{mod}}(x,\,t,\,\hbar)&\mbox{~if}\quad 1<x<C\hbar^{-\beta}
\\[3mm]
\chi_{Gauss}^\infty(x,t,\hbar)&\mbox{~if}\quad C\hbar^{-\beta}\leq x
\\[3mm]
0&\mbox{~otherwise}
\end{array}\ \right\}\quad\ +\quad
O\left(\,\hbar^{3/4+\delta}\,e^{-\,\alpha(E^*)/\hbar}\,\right),
$$
where the error term is measured in the $L^2$ norm,
$$
\chi_{{mod}}(x,\,t,\,\hbar)\ =\
\frac{P_{0}(x,E^*)\ \sqrt{2\,\pi\,\hbar\,}}
{\sqrt{\alpha''(E^*)\,+\,i\,S''(x,\,t,\,E^*)}}\
e^{-\,(\alpha(E^*)\,+\,i\,S(x,\,t,\,E^*))/\hbar}\
e^{-\,\frac{S'(x,\,t,\,E^*)^{2}}
{2\,\hbar\,(\alpha''(E^*)\,+\,i\,S''(x,\,t,\,E^*))}},
$$
and $\phantom{T}'$ denotes the derivative with respect to $E$.
\end{thm}

\vskip 5mm\noindent
{\bf Proof Outline}\quad
We follow the main steps of the proof of the corresponding result
for Theorem 5 of \cite{bjt09}, with one notable exception.
Since we do not use any superadiabatic representation,
the next to leading order term in the asymptotics of $c^a(x,\,E,\,\hbar)$
is of too low an order to be treated as in \cite{bjt09}.
We briefly address this issue in more detail.
By integration by parts, we see that for~ $x>1$,
$$
c^a(x,E,\hbar)\ =\
1\ +\ {i\,\hbar}\ \int_x^\infty\,
\frac{\left(\frac{\partial p}{\partial x}(y,\,E)\right)^2}{8\ p^3(y,\,E)}\ dy\
\,+\ \,
O_E(\hbar^2/{x^{\nu+1}}).
$$
When we integrate against the energy density $Q(E,\,\hbar)$,
the remainder term can be dealt with by using Lemma 1 of \cite{bjt09}.
The non-zero next to leading term is only of order $\hbar/{x^{\nu+1}}$
and the error term it generates can be bounded as follows:

Let~
$\eta(x,\,t,\,\hbar)\,=\,\chi(x,\,t,\,\hbar)\,-\,
\widetilde\chi(x,\,t,\,\hbar)$,~ where
\bea
\label{chitilde}
&&\widetilde\chi(x,\,E,\,\hbar)
\\[3mm]\nonumber
&=&\quad
\frac{e^{-K(E)/(2\hbar)}\ \,\sqrt{p(-\infty,\,E)\,}\ \,
e^{i\int_{x_1(E)}^x\,p_R(y,\,E)\,dy/\hbar}}
{\sqrt{p(x,\,E)\,}\ c_2^b(-\infty,\,E,\,\hbar)\
e^{-i\int_{-\infty}^{x_0(E)}(p_R(y,E)-p_R(-\infty,E))/\hbar}\
e^{-\,i\,p_R(-\infty,\,E)\,x_0(E)}}\,.
\eea
The error term whose $L^2$ norm we need to bound has the explicit form
\bea\nonumber
g(x,\,t,\,\hbar)&:=&
{i\,\hbar}\ \int_\Delta\ Q(E,\,\hbar)\ e^{-iEt/\hbar}\
\widetilde\chi(x,\,E,\,\hbar)\
\int_x^\infty\ 
\frac{\left(\frac{\partial p}{\partial x}(y,\,E)\right)^2}{8\ p^3(y,\,E)}\
dy\ dE
\\[3mm]\nonumber
&\equiv& \hbar\ \int_\Delta\
e^{-iEt/\hbar}\ \widetilde Q(E,\,\hbar)\ f(x,\,E)\
e^{i\int_{x_1(E)}^x\,p_R(y,\,E)\,dy/\hbar}\ dE,
\eea
where~ $f(x,\,E)=O(1/{x^{\nu+1}})$,~ uniformly for~ $E\in \Delta$,~
$\widetilde Q(E,\,\hbar)$~ is independent of~ $x$,
and~ $|\widetilde Q(E,\,\hbar)|$~ behaves essentially as~
$e^{-\alpha(E^*)/\hbar}$~
times a Gaussian in $(E-E^*)/\sqrt\hbar$. (See (\ref{gabe})).
We compute
\bea\label{g2}
\hspace{-14mm}&&\int_{x>1}\ |g(x,\,t,\,\hbar)|^2\ dx
\\[3mm]\nonumber
\hspace{-14mm}&=&\hbar^2\ \int_{\Delta\times \Delta}\,
\widetilde Q(E,\,\hbar)\,\overline{\widetilde Q}(E',\,\hbar)
\int_1^\infty f(x,\,E)\,\overline f(x,\,E')\,
e^{i\int_{x_1(E)}^x\,(p_R(y,\,E)-p_R(y,\,E'))\,dy/\hbar}\,dx\,dE\,dE'.
\eea
Let $0<\theta<1$.
We split the integration range into the sets where
$|E-E'|<\hbar^\theta$ and $|E-E'|\geq \hbar^\theta$.~
We perform an integration by parts in $x$
on the latter set to get
\bea\nonumber
\hspace{-15mm}&&\int_1^\infty\ f(x,\,E)\ \overline f(x,\,E')\
e^{i\int_{x_1(E)}^x\,(p_R(y,\,E)-p_R(y,\,E'))\,dy/\hbar}\,dx
\\[3mm] \nonumber
\hspace{-15mm}&=&\frac{i\,\hbar}{2\,(E-E')}\
(p_R(1,\,E)+p_R(1,\,E'))\ f(1,\,E)\
\overline f(1,\,E')\
e^{i\int_{x_1(E)}^1\ (p_R(y,\,E)-p_R(y,\,E'))\,dy/\hbar}
\\[3mm]\nonumber
\hspace{-15mm}&&+\quad\frac{i\,\hbar}{2(E-E')}
\int_1^\infty \frac{\partial\phantom{i}}{\partial x}
\left(\,{(p_R(x,E)+p_R(x,E'))}\ f(x,\,E)\ \overline f(x,\,E')\,\right)
\\[3mm]\nonumber
&&\hspace{8cm}
\times\quad e^{i\int_{x_1(E)}^x\,(p_R(y,\,E)-p_R(y,\,E'))\,dy/\hbar}\,dx.
\eea
The absolute values of both terms are bounded by~ $C\hbar^{1-\theta}$,~
where $C$ is uniform in $E$.
So, they contribute a factor $C\hbar^{3-\theta}\|\widetilde Q\|_{1}^2$
to (\ref{g2}).
Similarly, the integral in $x$ in (\ref{g2}) is bounded uniformly in
$E$ and, for some constant $C$,
$$
\int_{\Delta\times \Delta}\ |\widetilde Q(E,\hbar)|\
|\widetilde Q(E',\hbar)|\ \chi_{\{|E-E'|<\hbar^\theta\}}\ dE\ dE'\ \leq\
C\ \|\widetilde Q\|_{2}^2\ \hbar^{\theta/2}.
$$
Since $\Delta$ is compact,
$\|\widetilde Q \|_{1}\leq |\Delta |^{1/2} \|\widetilde Q \|_{2}$.
Using this and the the estimate\\
$\|\widetilde Q \|_{2}\leq C e^{-\alpha(E^*)/\hbar}\hbar^{1/4}$,~
we eventually get for $\theta=2/3$,
$$
\|g\|_2\ \le\
C\ e^{-\alpha(E^*)/\hbar}\ \hbar^{17/12}\ \ll\
C\ e^{-\alpha(E^*)/\hbar}\ \hbar^{3/4}.
$$

The rest of the proof, which consists of showing that $\widetilde\chi$
can be approximated by $\chi_{mod}$ and $\chi^\infty_{Gauss}$
for different values of $x$, now relies on Lemma \ref{23}
and on arguments identical to those in the proof
of Theorem 5 of \cite{bjt09}.
\hfill\ep

\vskip 5mm
While explicit and concise, the approximation above does not make
apparent where the transmitted wave is actually located.
To have a better idea of the position of this wave function,
we define $q_t$ to be the unique solution in $x$ to
$\ds\frac{\partial\phantom{i}}{\partial E}\,S(x,\,t,\,E^*)=0$
with $q_t>0$ and $\dot q_t >0$.
The function $q_t$ is actually the classical trajectory
in the potential $V$ with energy $E^*$,
the velocity of which is bounded from above and below.
We define
\bea\nonumber
\chi_{Gauss}(x,\,t,\,\hbar)&=&
\frac{P_{0}(q_t,\,E^*)\ \sqrt{2\,\pi\,\hbar\,}}
{\sqrt{\alpha''(E^*)\,+\,i\,S''(q_t,\,t,\,E^*)\,}}\ \,
\exp\left\{\,-\,(\alpha(E^*)\,+\,i\,S(x,\,t,\,E^*))/\hbar\,\right\}
\\[4mm]\nonumber
&&\times\quad
\exp\left\{\,-\ \frac{(x\,-\,q_t)^2}
{2\,\hbar\,p_R(q_t,\,E^*)^2\,
(\alpha''(E^*)\,+\,i\,S''(q_t,\,t,\,E^*))}\,\right\}.
\eea
This wave packet is a Gaussian
that is centered on the trajectory $q_t$,
and whose width is of order $\sqrt{\hbar}$.

This leads immediately to the following corollary.
(See Theorem 6 of \cite{bjt09}.)
%
%
\begin{cor}\label{growl}\quad
There exist~ $X_0>0$~ and~ $\delta>0$,~
such that for all times~ $t$,~ with\\ $X_0<q_t<C\hbar^{-\beta}$,~
we have, in the $L^2$ sense,
$$
\chi(x,\,t,\,\hbar)\ =\
\chi_{Gauss}(x,\,t,\,\hbar)\ +\
O\left(\,\hbar^{3/4+\delta}\ e^{-\alpha(E^*)/\hbar}\,\right),
$$
where\qquad
$\ds
\|\chi_{Gauss}(x,\,t,\,\hbar)\|_{L^2}
\ =\ O\left(\,\hbar^{3/4}\ e^{-\alpha(E^*)/\hbar}\,\right)
$.
\end{cor}

\end{document}